\PassOptionsToPackage{unicode}{hyperref}
\PassOptionsToPackage{hyphens}{url}
\PassOptionsToPackage{dvipsnames,svgnames,x11names}{xcolor}
\documentclass[
]{article}
\usepackage{xcolor}
\usepackage[margin=1in]{geometry}
\usepackage{amsmath,amssymb}
\usepackage{iftex}
\ifPDFTeX
  \usepackage[T1]{fontenc}
  \usepackage[utf8]{inputenc}
  \usepackage{textcomp} 
\else 
  \usepackage{unicode-math} 
  \defaultfontfeatures{Scale=MatchLowercase}
  \defaultfontfeatures[\rmfamily]{Ligatures=TeX,Scale=1}
\fi
\usepackage{lmodern}
\ifPDFTeX\else
\fi
\IfFileExists{upquote.sty}{\usepackage{upquote}}{}
\IfFileExists{microtype.sty}{
  \usepackage[]{microtype}
  \UseMicrotypeSet[protrusion]{basicmath} 
}{}
\makeatletter
\@ifundefined{KOMAClassName}{
  \IfFileExists{parskip.sty}{%
    \usepackage{parskip}
  }{
    \setlength{\parindent}{0pt}
    \setlength{\parskip}{6pt plus 2pt minus 1pt}}
}{
  \KOMAoptions{parskip=half}}
\makeatother
\usepackage{color}
\usepackage{fancyvrb}

\DefineVerbatimEnvironment{Highlighting}{Verbatim}{commandchars=\\\{\}}
\usepackage{framed}
\definecolor{shadecolor}{RGB}{248,248,248}
\newenvironment{Shaded}{\begin{snugshade}}{\end{snugshade}}

\newcommand{\CharTok}[1]{\textcolor[rgb]{0.31,0.60,0.02}{#1}}
\newcommand{\CommentTok}[1]{\textcolor[rgb]{0.56,0.35,0.01}{\textit{#1}}}

\newcommand{\ControlFlowTok}[1]{\textcolor[rgb]{0.13,0.29,0.53}{\textbf{#1}}}
\newcommand{\DataTypeTok}[1]{\textcolor[rgb]{0.13,0.29,0.53}{#1}}
\newcommand{\DecValTok}[1]{\textcolor[rgb]{0.00,0.00,0.81}{#1}}

\newcommand{\KeywordTok}[1]{\textcolor[rgb]{0.13,0.29,0.53}{\textbf{#1}}}
\newcommand{\NormalTok}[1]{#1}
\newcommand{\OperatorTok}[1]{\textcolor[rgb]{0.81,0.36,0.00}{\textbf{#1}}}

\newcommand{\SpecialCharTok}[1]{\textcolor[rgb]{0.81,0.36,0.00}{\textbf{#1}}}

\newcommand{\StringTok}[1]{\textcolor[rgb]{0.31,0.60,0.02}{#1}}

\usepackage{longtable,booktabs,array}
\usepackage{calc} 
\usepackage{etoolbox}
\makeatletter
\patchcmd\longtable{\par}{\if@noskipsec\mbox{}\fi\par}{}{}
\makeatother
\IfFileExists{footnotehyper.sty}{\usepackage{footnotehyper}}{\usepackage{footnote}}
\makesavenoteenv{longtable}
\usepackage{graphicx}
\makeatletter
\newsavebox\pandoc@box
\newcommand*\pandocbounded[1]{
  \sbox\pandoc@box{#1}%
  \Gscale@div\@tempa{\textheight}{\dimexpr\ht\pandoc@box+\dp\pandoc@box\relax}%
  \Gscale@div\@tempb{\linewidth}{\wd\pandoc@box}%
  \ifdim\@tempb\p@<\@tempa\p@\let\@tempa\@tempb\fi
  \ifdim\@tempa\p@<\p@\scalebox{\@tempa}{\usebox\pandoc@box}%
  \else\usebox{\pandoc@box}%
  \fi%
}
\def\fps@figure{htbp}
\makeatother
\providecommand{\tightlist}{%
  \setlength{\itemsep}{0pt}\setlength{\parskip}{0pt}}
\usepackage{pdflscape}
\usepackage{bookmark}
\IfFileExists{xurl.sty}{\usepackage{xurl}}{} 
\hypersetup{
  pdftitle={SMS-delivered network-initiated SUPL on Pixel 8: a privacy assessment},
  colorlinks=true,
  linkcolor={Maroon},
  filecolor={Maroon},
  citecolor={Blue},
  urlcolor={Blue},
  pdfcreator={LaTeX via pandoc}}

\title{SMS-delivered network-initiated SUPL on Pixel 8: a privacy
assessment}
\author{Douglas Leith\\
Trinity College Dublin, Ireland\\
doug.leith@tcd.ie\\
19 October 2026}
\date{}

\begin{document}
\maketitle

\subsection{1. Summary}\label{summary}

A SUPL\_INIT message is a network-initiated trigger that can be sent to
a handset using an SMS to unilaterally start a location session: on
receipt, the handset is instructed to determine its own position and
report it, together with an identifier such as its IMSI, to a server
specified in the message, without any action by the phone's user. The
concern motivating this investigation is therefore whether such a
message could be used to silently exfiltrate a handset's location and
subscriber identity to a server under an attacker's control, with no
consent from or visibility to the device's owner.

We investigated this on a Google Pixel 8 handset, which uses a Samsung
Exynos modem and a Broadcom GPS/GNSS subsystem; the conclusions below
are specific to this combination of modem and GNSS vendor and may not
apply to phones built around different components. On this device, an
SMS-delivered SUPL\_INIT is received by the Samsung modem, but
essentially all of the SUPL-specific processing (every check on whether
and how to act on the message) takes place in the Broadcom GNSS vendor
code, principally the \texttt{gpsd} daemon; the modem's role is limited
to passing the raw message on to \texttt{gpsd}.

We find no privacy issue: the handset never sends location data to an
attacker-chosen server as a result of an unsolicited SUPL\_INIT
delivered by SMS.

This follows from three checks and behaviours we identify in
\texttt{gpsd}, established by static analysis of the decompiled binary
and confirmed by dynamic, live testing with a real SMS delivered over
the air:

\begin{enumerate}
\def\labelenumi{\arabic{enumi}.}
\tightlist
\item
  \textbf{A SUPL\_INIT packet is ignored unless the handset is already
  in an active emergency call.} The response to a network-initiated
  SUPL\_INIT gates on the handset's own emergency-call state at the time
  the message arrives, not on any claim made by the packet itself. An
  attacker sending an SMS cannot place the handset into a real emergency
  call, so this check blocks every SUPL\_INIT an attacker could actually
  deliver.
\item
  \textbf{When the handset is already in an active emergency call, and
  the received SUPL\_INIT is itself flagged emergency with a valid SLP
  address, that address is used.} The code supports restricting this to
  APNs on a configured allowlist, but on this device that allowlist is
  empty (unconfigured in \texttt{gps.xml}), so the check is skipped and
  does not currently restrict anything. This is the one path in the code
  by which an attacker-supplied server address could be used at all,
  and, regardless of the allowlist, it requires a precondition (a live
  emergency call) that is outside an SMS attacker's control.
\item
  \textbf{For a non-emergency SUPL\_INIT, the server address carried in
  the message is ignored.} This is the message any purely SMS-based
  attack must send, since it cannot place the handset into a real
  emergency call. The destination is always the handset's own configured
  SLP server (or a standard, SIM-derived hostname if that is unset),
  never an address supplied in the packet: the attacker cannot control
  where location data is sent.
\end{enumerate}

Separately, and independently of any network-initiated SUPL\_INIT
message, the handset also generates its own SUPL requests to
\texttt{supl.google.com} as part of normal operation (background
assisted-GPS synchronisation). These internally generated requests are
not affected by network-initiated SUPL\_INIT messages.

We also compare the Pixel 8 with an iPhone 8 (Section 6). On the iPhone
studied, SUPL is disabled: the handset ignores any SUPL\_INIT message it
receives and does not open SUPL connections of its own. More generally,
iOS acts on a SUPL\_INIT only during an emergency call (no carrier
bundle relaxes this), and connects only to a server that matches the
carrier's configured one.

\subsection{2. SUPL in the public
specifications}\label{supl-in-the-public-specifications}

This section summarises what the publicly available Open Mobile Alliance
(OMA) specifications say about SUPL, SUPL\_INIT, its delivery mechanism
and its authentication requirements, independently of anything found by
static or dynamic analysis of this specific handset. We draw on the SUPL
2.0 Architecture Document {[}OMA-AD-SUPL{]} and the SUPL 2.0 User Plane
Location Protocol technical specification {[}OMA-TS-ULP{]} (full
references in Section 8).

\subsubsection{2.1 Purpose of SUPL}\label{purpose-of-supl}

Secure User Plane Location (SUPL) is an OMA enabler that carries
assistance data and positioning data over an IP (user-plane) bearer ``to
aid network and SET based positioning technologies in the calculation of
a SET's position'' {[}OMA-AD-SUPL, §5{]} (``SET'', SUPL Enabled
Terminal, is the specification's term for the handset). It exists as an
alternative to legacy control-plane positioning, using ordinary IP
connectivity rather than signalling channels, and defines both
device-initiated flows (the handset requests its own position) and
\textbf{network-initiated flows}, in which the network-side server, an
SLP (SUPL Location Platform), unilaterally starts a location session
with a handset.

\subsubsection{2.2 What SUPL\_INIT is, and how it reaches the
handset}\label{what-supl_init-is-and-how-it-reaches-the-handset}

SUPL\_INIT is the message used to start a network-initiated session,
sent by the Home SUPL Location Platform (H-SLP, the SLP belonging to the
SET's home network): in every network-initiated call flow in the
specification, ``the H-SLP initiates the SUPL Session with the SET by
sending a ULP SUPL INIT message'' {[}OMA-TS-ULP, §6, call-flow steps{]}.
Its ASN.1 definition carries, at minimum, a session identifier and the
intended positioning method, plus (when the network's own privacy check
decides notification or verification is needed) a \texttt{Notification}
element:

\begin{verbatim}
SUPLINIT ::= SEQUENCE {
  posMethod    PosMethod,
  notification Notification OPTIONAL,
  sLPAddress   SLPAddress OPTIONAL,
  qoP          QoP OPTIONAL,
  sLPMode      SLPMode,
  mac          MAC OPTIONAL,          -- backwards compatibility
  keyIdentity  KeyIdentity OPTIONAL,  -- backwards compatibility
  ...,
  ver2-SUPL-INIT-extension Ver2-SUPL-INIT-extension OPTIONAL }

NotificationType ::= ENUMERATED {
  noNotificationNoVerification(0), notificationOnly(1),
  notificationAndVerficationAllowedNA(2),
  notificationAndVerficationDeniedNA(3), privacyOverride(4), ...}
\end{verbatim}

{[}OMA-TS-ULP, §11{]}

\texttt{privacyOverride} allows the location session to proceed with no
notification shown to the user at all: this is the notification type
used throughout this investigation's test payloads.

The specification defines four transport mechanisms for delivering
SUPL\_INIT to the handset: OMA Push (WAP Push), SMS directly, UDP/IP,
and SIP Push. It mandates that ``For GSM/WCDMA/TD-SCDMA deployments, the
SIF {[}SUPL Initiation Function{]} using OMA Push SHALL be supported by
both the SET and the SLP'' {[}OMA-AD-SUPL, §5.3.1.2{]}.
WAP-Push-over-SMS delivery, as used throughout this investigation, is
therefore a mandatory transport for the class of network the test
handset operates on, not an implementation quirk of this particular
device.

The handset's own response to a SUPL\_INIT is one of
\texttt{SUPL\ POS\ INIT}, \texttt{SUPL\ AUTH\ REQ} or
\texttt{SUPL\ TRIGGERED\ START} {[}OMA-TS-ULP, §6.1.6.1{]}, and a
session normally concludes with a \texttt{SUPL\ END} message. Both
\texttt{SUPLPOSINIT} and \texttt{SUPLEND} carry an optional
\texttt{position} field {[}OMA-TS-ULP, §11{]}, and every message in the
session, including these, carries a \texttt{SetSessionID} identifying
which handset it belongs to. The specification defines this identifier
as explicitly allowed to be the handset's own IMSI:

\begin{verbatim}
SetSessionID ::= SEQUENCE {sessionId INTEGER(0..65535), setId SETId}

SETId ::= CHOICE {
  msisdn OCTET STRING(SIZE (8)), mdn OCTET STRING(SIZE (8)),
  min BIT STRING(SIZE (34)), imsi OCTET STRING(SIZE (8)),
  nai IA5String(SIZE (1..1000)), iPAddress IPAddress, ...,
  ver2-imei OCTET STRING(SIZE(8))}
\end{verbatim}

{[}OMA-TS-ULP, §11{]}

So the handset's response to a SUPL\_INIT is designed to deliver both
its own computed position and a uniquely identifying value (IMSI, MSISDN
or IMEI) to whichever server the session was directed to. This is the
basis for the concern in Section 1: it is not merely that the handset
computes its own position, but that its response is designed to carry
that position together with a unique subscriber or device identifier, to
a server the network message itself specifies.

\subsubsection{2.3 Authentication and verification mandated for
SUPL\_INIT}\label{authentication-and-verification-mandated-for-supl_init}

The specification's own security model for SUPL\_INIT is markedly weaker
than for the rest of the protocol: ``All SUPL Messages except `SUPL
INIT' MUST be delivered within a TLS or PSK-TLS session between a SET
and an SLP'' {[}OMA-AD-SUPL, §4.2.3{]}. SUPL\_INIT itself is explicitly
excluded, since it is the trigger that precedes any session or TLS
connection.

Two independent protections are defined for SUPL\_INIT {[}OMA-TS-ULP,
§6.1.6{]}:

\begin{itemize}
\tightlist
\item
  \textbf{Network-based authentication} (mandatory): the \emph{first
  message the handset sends back} after acting on a SUPL\_INIT
  (\texttt{SUPL\ POS\ INIT}, \texttt{SUPL\ AUTH\ REQ}, etc.) must carry
  a verification field computed as an HMAC over the received SUPL\_INIT;
  if this fails, the SLP terminates the session. This is a check the
  \emph{network} performs on the handset's response, \textbf{after} the
  handset has already processed the SUPL\_INIT and acted on it; it
  cannot itself prevent the handset acting on a forged message.
\item
  \textbf{End-to-end protection} (optional): a
  \texttt{Protection\ Level} field in SUPL\_INIT is either \texttt{Null}
  (``no end-to-end integrity protection, no end-to-end replay protection
  and no confidentiality protection'') or \texttt{Basic} (a 32-bit
  HMAC-SHA256 MAC plus a replay counter, keyed from a key established
  during a prior GBA/SEK-authenticated TLS session with the home SLP)
  {[}OMA-TS-ULP, §6.1.6.3, §6.1.6.6{]}. Critically, \texttt{Null}
  protection is explicitly spec-compliant, is the assigned default ``at
  power-up or when the lifetime of the SUPL\_INIT\_Root\_Key has
  expired'' {[}§6.1.6.4{]}, and under it ``the SET considers the
  {[}received{]} message to be authentic, and no security related
  processing is required'' {[}§6.1.6.5{]}. That is, by design, a handset
  that has not recently negotiated \texttt{Basic} protection with its
  home SLP accepts any correctly-tagged SUPL\_INIT as authentic, with no
  cryptographic check at all.
\end{itemize}

In short: the public specification's own authentication model for
SUPL\_INIT is optional, falls back to no protection whenever a prior
key-establishment session has not occurred, and even when active only
protects against a network guessing wrong, not a third party
constructing a well-formed message. The behaviour we found on this
device (gating on the handset's own emergency-call state) is an
Android/vendor-level defence that goes beyond what SUPL 2.0 itself
requires; it is not something the specification mandates.

A later version, SUPL 3.0 {[}OMA-TS-ULP-V3{]}, adds a third protection
mode but does not change this picture. \texttt{Null} protection (no
end-to-end protection at all) remains explicitly listed as ``Optional''
and stays the mandatory fallback whenever no key has been established,
and support for the newer \texttt{Mode\ A} protection is only
recommended (``SHOULD''), not required, for a SET to implement at all
{[}OMA-TS-ULP-V3, §6.3.1, Table 5{]}. Its emergency exemption is, if
anything, stronger than SUPL 2.0's: ``During an emergency call, a SET
SHALL NOT apply end-to-end protection of emergency SUPL INIT messages''
{[}OMA-TS-ULP-V3, §6.2.4{]}. The test handset's own configuration
declares \texttt{SuplVersion="2"}, \texttt{SuplMinorVersion="0"}
(\texttt{gps.xml}), so it runs SUPL 2.0 rather than 3.0; this paragraph
is included for completeness, not because it changes any finding in this
report.

\subsubsection{2.4 SUPL\_INIT for emergency
calls}\label{supl_init-for-emergency-calls}

The specification treats emergency positioning as a distinct case with,
if anything, \emph{weaker} message-level protection than ordinary
sessions: ``End-to-End Protection of SUPL INIT Messages applies only to
non-emergency SUPL INIT messages'' {[}OMA-TS-ULP, §6.1.6.3{]}. An
emergency SUPL\_INIT can only ever rely on the after-the-fact
network-based check above. The specification also allows a handset with
no SIM/UICC at all to use SUPL for emergency positioning, authenticated
only weakly ``using (e.g.) the session ID and the received hash of the
SUPL INIT'' {[}§6.1.5.4{]}, and provides a mechanism for an emergency
SLP to resend a corrected SUPL\_INIT over an established TLS session if
it detects the original was altered, explicitly noting ``the ability to
resend SUPL INIT is only intended for emergency sessions''
{[}§6.1.5.5{]}.

Beyond these provisions, the public documents give little detail on how
an individual network operator or device vendor actually gates emergency
SUPL\_INIT processing in practice: for example, on what signal a handset
uses to decide it is ``in an emergency call'' for this purpose, or what
(if any) allowlisting of destination addresses is applied. The
Architecture Document notes only that the network's privacy function
should ``allow override of the target SET User privacy settings as
mandated or allowed by local regulations for positioning for an
emergency services call'' {[}OMA-AD-SUPL, §5.3.1.1{]}. That is, it
delegates the actual policy to regulation and implementation, rather
than specifying it. The emergency-call-state gate and APN allowlist
described in Section 1 (findings 1--2) are exactly this kind of
implementation-specific policy, found by analysing this handset's own
code rather than documented in the public SUPL specifications.

\subsection{3. Experimental setup}\label{experimental-setup}

\textbf{Handset.} Google Pixel 8 (``shiba''), Samsung Exynos Modem 5300
baseband, Broadcom BCM4776-family GNSS RF front-end (\texttt{RfType} is
configured as \texttt{GL\_RF\_4776\_BRCM} in \texttt{gps.xml}), Android
14 (build \texttt{AP2A.240905.003}, security patch level 5 September
2024). Rooted with Magisk. All testing was performed by the researcher
sending SMS messages to their own device.

\textbf{Static analysis.} The vendor GNSS daemon binary
(\texttt{/vendor/bin/hw/gpsd}, Broadcom's proprietary \texttt{glw}/GPS
HAL library, statically linked, stripped but with recoverable C++ symbol
names) was decompiled with Ghidra 11.4.2. The binary self-identifies as
\texttt{Broadcom\ GLL\ ver.\ 154.20.24} (build 587347, build\_job\_id
477614240, dated 12 January 2024), built against Google's ``P23''
Android target.

\textbf{Dynamic analysis.} Frida 17.18.0 was used to instrument the
running \texttt{gpsd} process on-device. Instrumentation attached to the
already-running, normally \texttt{init}-started \texttt{gpsd} process.

\textbf{SMS trigger.} A minimal Android app (\texttt{apk/}, package
\texttt{com.test.suplsms}) sends one binary/port SMS
(\texttt{SmsManager.sendDataMessage}, port 2948) containing a
WAP-push-wrapped SUPL\_INIT to the device's own number. Both parts of
this trigger are fully defined in the public specification: the
ULP-PDU/SUPL\_INIT ASN.1 grammar in {[}OMA-TS-ULP, §11{]}, and the exact
WSP/WAP-Push OTA byte layout (PDU type, content-type value
\texttt{0x0312} \texttt{application/vnd.omaloc-supl-init}, and the
OMNA-registered application id \texttt{x-oma-application:ulp.ua}) as a
fully worked, byte-level example in {[}OMA-TS-ULP, Annex B.2--B.3, Table
82{]}. The ULP-PDU payload itself was built with Python's
\texttt{asn1tools} (0.169.0) against this ASN.1 schema
(\texttt{payload\_builder/ulp.asn}), producing genuine,
standards-conformant SUPL\_INIT messages. The trigger payload set
\texttt{posMethod} to \texttt{agpsSETassisted},
\texttt{notificationType} to \texttt{privacyOverride} (the
no-notification, no-verification case described in Section 2.3), and
\texttt{sLPAddress} to the IPv4 address of the test server (Section 3,
\texttt{Test\ server}, encoded as an \texttt{iPAddress} choice rather
than a \texttt{fQDN}), with \texttt{sLPMode} set to \texttt{nonProxy}.
The public specification, on its own, was sufficient to construct a
byte-correct trigger SMS; no analysis of Android or vendor code was
needed for this part.

\textbf{Test server.} For dynamic testing of the
connection-establishment logic, a Python TLS server was built presenting
a leaf certificate signed by a CA installed as a trusted system CA on
the test handset.

\subsection{4. Static analysis}\label{static-analysis}

\subsubsection{4.1 System-level message
flow}\label{system-level-message-flow}

An SMS-delivered SUPL\_INIT travels from the carrier network, through
the modem baseband and Android's RIL layer, into \texttt{gpsd}. If
\texttt{gpsd} decides to act on it, the message goes on via
inter-process communication (IPC) to \texttt{scd}, the separate daemon
that owns the actual outbound network socket.

\begin{figure}
\centering
\includegraphics[width=1\linewidth,height=\textheight,keepaspectratio,alt={System-level flow of a network-initiated SUPL message from SMS delivery to the outbound network connection. gpsd and scd are separate processes communicating over a named-pipe IPC channel (/data/vendor/gps/.pipe.gpsd\_to\_scd.*); gpsd decides whether and where to connect, scd owns the socket that actually does so.}]{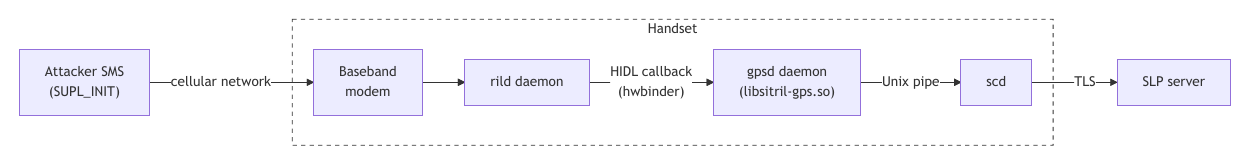}
\caption{System-level flow of a network-initiated SUPL message from SMS
delivery to the outbound network connection. \protect\texttt{gpsd} and
\protect\texttt{scd} are separate processes communicating over a
named-pipe IPC channel
(\protect\texttt{/data/vendor/gps/.pipe.gpsd\_to\_scd.*});
\protect\texttt{gpsd} decides whether and where to connect,
\protect\texttt{scd} owns the socket that actually does
so.\label{fig:system-flow}}
\end{figure}

\subsubsection{\texorpdfstring{4.2 Execution flow and key checks inside
\texttt{gpsd}}{4.2 Execution flow and key checks inside gpsd}}\label{execution-flow-and-key-checks-inside-gpsd}

Figure 2 shows the execution flow inside \texttt{gpsd} from SUPL\_INIT
message to connection request, with the two decision points that
determine outcome: the emergency-APN allowlist check (node E, only
relevant to emergency-flagged packets) and the NFW permission gate (node
G), which applies unconditionally.

\begin{figure}[htbp]
\centering
\begin{minipage}[b]{0.3653\linewidth}
\centering
\includegraphics[width=\linewidth]{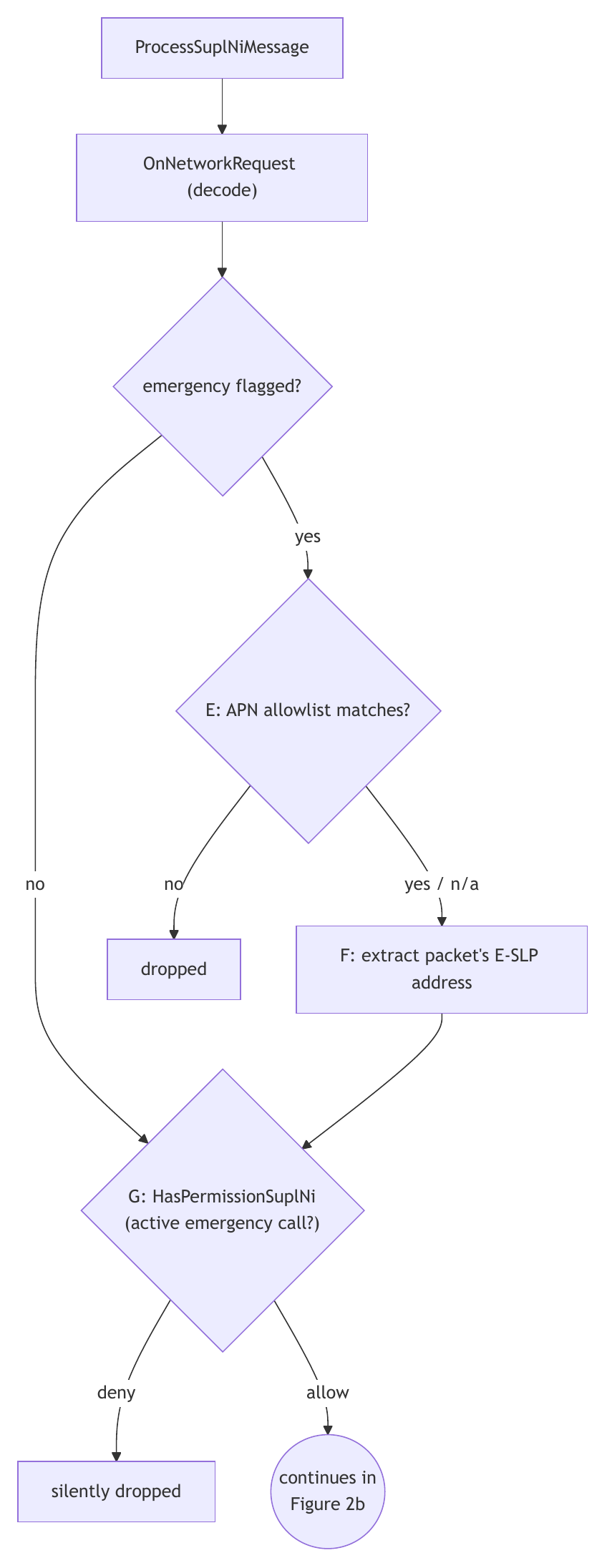}\\[2pt]
(a)
\end{minipage}\hfill
\begin{minipage}[b]{0.4056\linewidth}
\centering
\includegraphics[width=\linewidth]{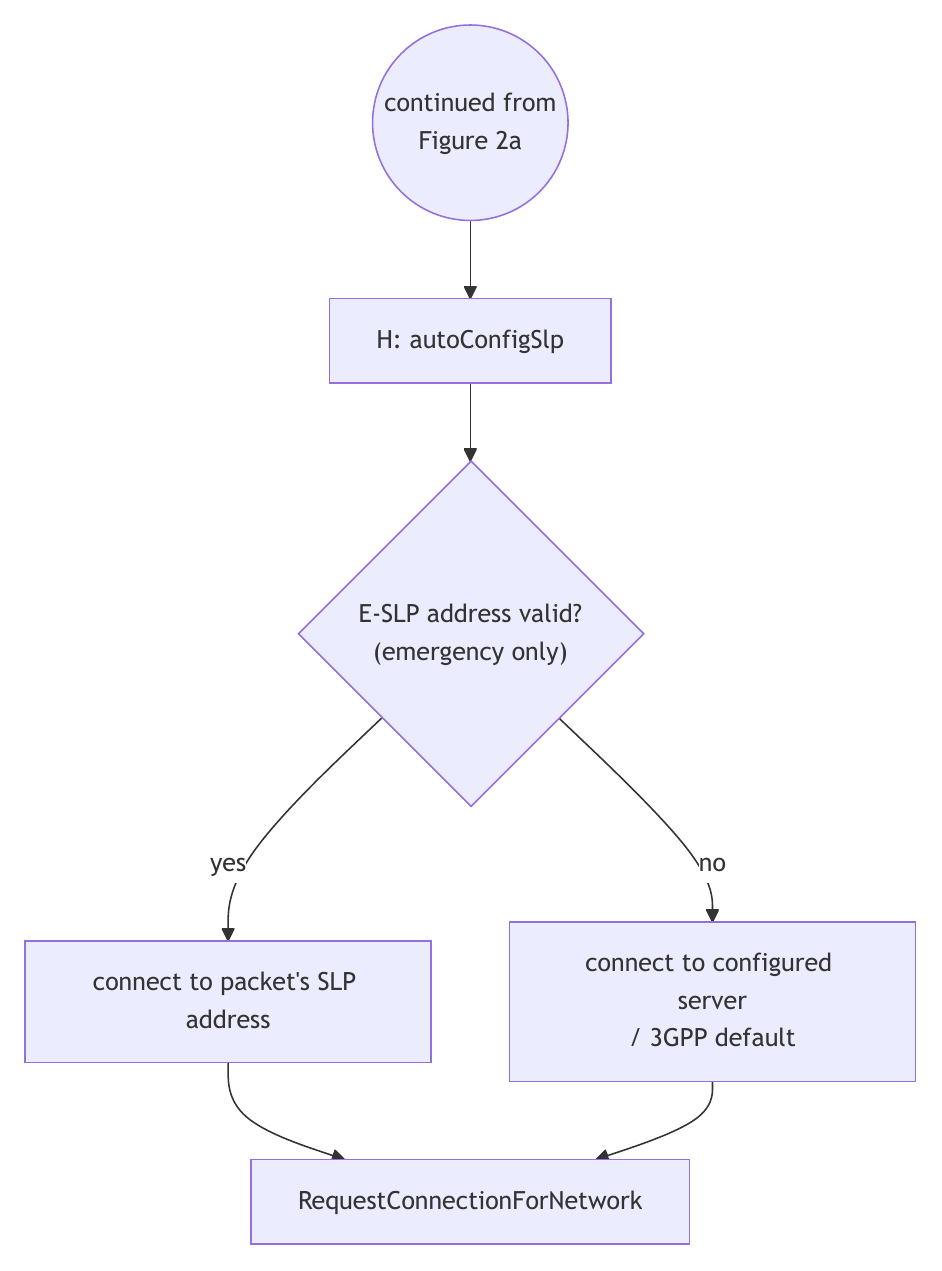}\\[2pt]
(b)
\end{minipage}
\caption{Execution flow inside \texttt{gpsd} from SUPL\_INIT message to
connection request, with the two decision points that determine
outcome: the emergency-APN allowlist check (node E, only relevant to
emergency-flagged packets) and the NFW permission gate (node G), which
applies unconditionally. The left part (2a) covers decode through the
permission gate; the right part (2b) continues from there through
destination resolution (node H) to the connection request. Function
names are as recovered from the binary's C++ symbols and can be located
directly in the decompiled output.}
\label{fig:gpsd-flow}
\end{figure}

\subsubsection{4.3 The permission gate}\label{the-permission-gate}

\texttt{RildClientHelper::HasPermissionSuplNi} is the check
corresponding to node G in Figure \ref{fig:gpsd-flow}. Decompiled
(cleaned up for readability; original at file offset \texttt{0x339640}):

\begin{Shaded}
\begin{Highlighting}[]
\DataTypeTok{bool}\NormalTok{ RildClientHelper}\OperatorTok{::}\NormalTok{HasPermissionSuplNi}\OperatorTok{(}
\NormalTok{        RildClientHelper }\OperatorTok{*}\NormalTok{this}\OperatorTok{,} \DataTypeTok{bool}\NormalTok{ isEmergencySuplNi}\OperatorTok{,} \DataTypeTok{bool}\NormalTok{ isInEmergencyState}\OperatorTok{)}
\OperatorTok{\{}
  \DataTypeTok{char}\NormalTok{ cVar1 }\OperatorTok{=} \OperatorTok{*}\NormalTok{SuplIgnoreNfwLocPolicy}\OperatorTok{;}   \CommentTok{// configuration property}
  \ControlFlowTok{if} \OperatorTok{(}\NormalTok{cVar1 }\OperatorTok{==} \CharTok{\textquotesingle{}}\SpecialCharTok{\textbackslash{}0}\CharTok{\textquotesingle{}} \OperatorTok{\&\&} \OperatorTok{!}\NormalTok{isInEmergencyState}\OperatorTok{)} \OperatorTok{\{}
\NormalTok{    ILog}\OperatorTok{::}\NormalTok{Log}\OperatorTok{(...,} \StringTok{"SUPL: SUPL NI is not allowed. attribution app {-} }\SpecialCharTok{\%s}\StringTok{"}\OperatorTok{,}
\NormalTok{              attributionAppPkgName}\OperatorTok{);}
  \OperatorTok{\}}
  \ControlFlowTok{return}\NormalTok{ cVar1 }\OperatorTok{!=} \CharTok{\textquotesingle{}}\SpecialCharTok{\textbackslash{}0}\CharTok{\textquotesingle{}} \OperatorTok{||}\NormalTok{ isInEmergencyState}\OperatorTok{;}
\OperatorTok{\}}
\end{Highlighting}
\end{Shaded}

The decision depends only on \texttt{isInEmergencyState} (whether the
handset itself is currently in an active emergency call, passed in from
the telephony stack, outside the SMS attacker's control) and a
configuration flag \texttt{SuplIgnoreNfwLocPolicy} that is
\texttt{false} on this device (\texttt{gps.xml}, no override). It does
\textbf{not} depend on \texttt{isEmergencySuplNi}: whether the
\emph{packet itself} claims to be an emergency SUPL\_INIT is irrelevant
to this check. This substantiates conclusion 1.

\subsubsection{4.4 The emergency-SLP extraction and APN
allowlist}\label{the-emergency-slp-extraction-and-apn-allowlist}

Node E/F in Figure \ref{fig:gpsd-flow}, inside
\texttt{GlSuplHalPlatform::OnNetworkRequest} (file offset
\texttt{0x3d1718}):

\begin{Shaded}
\begin{Highlighting}[]
\ControlFlowTok{if} \OperatorTok{(}\NormalTok{decoded\_ok }\OperatorTok{\&\&}\NormalTok{ packet\_is\_emergency\_notification}\OperatorTok{)} \OperatorTok{\{}
\NormalTok{    this}\OperatorTok{{-}\textgreater{}}\NormalTok{isEmergencySuplNi }\OperatorTok{=} \KeywordTok{true}\OperatorTok{;}
    \ControlFlowTok{if} \OperatorTok{(}\NormalTok{SuplConfig}\OperatorTok{::}\NormalTok{Instance}\OperatorTok{(){-}\textgreater{}}\NormalTok{emergencyApnList}\OperatorTok{[}\DecValTok{0}\OperatorTok{]} \OperatorTok{!=} \CharTok{\textquotesingle{}}\SpecialCharTok{\textbackslash{}0}\CharTok{\textquotesingle{}}\OperatorTok{)} \OperatorTok{\{}
        \ControlFlowTok{if} \OperatorTok{(!}\NormalTok{apn\_in\_list}\OperatorTok{(}\NormalTok{current\_apn}\OperatorTok{,}\NormalTok{ SuplConfig}\OperatorTok{::}\NormalTok{Instance}\OperatorTok{(){-}\textgreater{}}\NormalTok{emergencyApnList}\OperatorTok{))} \OperatorTok{\{}
\NormalTok{            ILog}\OperatorTok{::}\NormalTok{Log}\OperatorTok{(...,} \StringTok{"SUPL: Current APN }\SpecialCharTok{\textbackslash{}"\%s\textbackslash{}"}\StringTok{ does not match "}
                           \StringTok{"emergency APN list }\SpecialCharTok{\textbackslash{}"\%s\textbackslash{}"\textbackslash{}n}\StringTok{"}\OperatorTok{,}\NormalTok{ current\_apn}\OperatorTok{,} \OperatorTok{...);}
\NormalTok{            ILog}\OperatorTok{::}\NormalTok{Log}\OperatorTok{(...,} \StringTok{"SUPL: Emergency SUPL INIT will be dropped}\SpecialCharTok{\textbackslash{}n}\StringTok{"}\OperatorTok{);}
            \ControlFlowTok{goto}\NormalTok{ dropped}\OperatorTok{;}
        \OperatorTok{\}}
    \OperatorTok{\}}
    \ControlFlowTok{if} \OperatorTok{(}\NormalTok{packet\_has\_slp\_address}\OperatorTok{)} \OperatorTok{\{}
\NormalTok{        this}\OperatorTok{{-}\textgreater{}}\NormalTok{eslpHost }\OperatorTok{=}\NormalTok{ decode\_slp\_address}\OperatorTok{(...);}   \CommentTok{// this+0x89}
\NormalTok{        this}\OperatorTok{{-}\textgreater{}}\NormalTok{eslpPort }\OperatorTok{=}\NormalTok{ decode\_slp\_port\_or\_default}\OperatorTok{(...);}
\NormalTok{        ILog}\OperatorTok{::}\NormalTok{Log}\OperatorTok{(...,} \StringTok{"SUPL: E{-}SLP: }\SpecialCharTok{\%s}\StringTok{:}\SpecialCharTok{\%d\textbackslash{}n}\StringTok{"}\OperatorTok{,}\NormalTok{ this}\OperatorTok{{-}\textgreater{}}\NormalTok{eslpHost}\OperatorTok{,}\NormalTok{ this}\OperatorTok{{-}\textgreater{}}\NormalTok{eslpPort}\OperatorTok{);}
    \OperatorTok{\}}
\OperatorTok{\}}
\end{Highlighting}
\end{Shaded}

The packet's own SLP address is only extracted at all when the packet is
flagged emergency, and, if an emergency-APN allowlist is configured,
only when the handset's current APN matches an entry in it.

\textbf{On this device, the allowlist is not configured.}
\texttt{SuplConfig}'s constructor zero-initialises the field
(\texttt{*(undefined8\ *)(this\ +\ 0x577)\ =\ 0;}), and neither of the
two \texttt{gps.xml} attribute names \texttt{SetCfgValue} accepts for it
(\texttt{E911Apns}, \texttt{SuplEmergencyApns}) appears in this device's
\texttt{gps.xml}. The APN check is therefore inactive here: an
emergency-flagged SUPL\_INIT with a decoded SLP address, once past the
permission gate in Section 4.3, would have that address used
unconditionally, with no APN restriction.

This substantiates conclusion 2: an attacker-controlled destination
address is reachable through this code only for emergency-flagged
messages, subject to an APN allowlist that exists in the code but is not
actually configured on this device. Per Section 4.3, it also only ever
takes effect if the handset is already in a genuine active emergency
call.

\subsubsection{4.5 Destination-address
resolution}\label{destination-address-resolution}

Node H in Figure \ref{fig:gpsd-flow},
\texttt{GlSuplHalPlatform::autoConfigSlp} (file offset
\texttt{0x4cff08}):

\begin{Shaded}
\begin{Highlighting}[]
\DataTypeTok{void}\NormalTok{ GlSuplHalPlatform}\OperatorTok{::}\NormalTok{autoConfigSlp}\OperatorTok{(}\NormalTok{GlSuplHalPlatform }\OperatorTok{*}\NormalTok{this}\OperatorTok{,} \DataTypeTok{bool}\NormalTok{ force}\OperatorTok{)}
\OperatorTok{\{}
  \ControlFlowTok{if} \OperatorTok{(!}\NormalTok{force}\OperatorTok{)} \OperatorTok{\{}
    \ControlFlowTok{if} \OperatorTok{(}\NormalTok{this}\OperatorTok{{-}\textgreater{}}\NormalTok{isEmergencySuplNi}\OperatorTok{)} \OperatorTok{\{}
      \DataTypeTok{char} \OperatorTok{*}\NormalTok{addr }\OperatorTok{=}\NormalTok{ this}\OperatorTok{{-}\textgreater{}}\NormalTok{eslpHost}\OperatorTok{[}\DecValTok{0}\OperatorTok{]} \OperatorTok{?}\NormalTok{ this}\OperatorTok{{-}\textgreater{}}\NormalTok{eslpHost}
                                      \OperatorTok{:}\NormalTok{ SuplConfig}\OperatorTok{::}\NormalTok{Instance}\OperatorTok{(){-}\textgreater{}}\NormalTok{server}\OperatorTok{;}
\NormalTok{      this}\OperatorTok{{-}\textgreater{}}\NormalTok{connectHost }\OperatorTok{=}\NormalTok{ addr}\OperatorTok{;}
\NormalTok{      this}\OperatorTok{{-}\textgreater{}}\NormalTok{connectPort }\OperatorTok{=}\NormalTok{ eslp\_port\_or\_configured\_default}\OperatorTok{();}
      \ControlFlowTok{if} \OperatorTok{(}\NormalTok{strcmp}\OperatorTok{(}\NormalTok{addr}\OperatorTok{,} \StringTok{"none"}\OperatorTok{)} \OperatorTok{!=} \DecValTok{0} \OperatorTok{\&\&}\NormalTok{ strcmp}\OperatorTok{(}\NormalTok{addr}\OperatorTok{,} \StringTok{"auto"}\OperatorTok{)} \OperatorTok{!=} \DecValTok{0}\OperatorTok{)}
        \ControlFlowTok{goto}\NormalTok{ use\_address}\OperatorTok{;}
    \OperatorTok{\}}
    \ControlFlowTok{if} \OperatorTok{(!}\NormalTok{server\_is}\OperatorTok{(}\StringTok{"none"}\OperatorTok{)} \OperatorTok{\&\&} \OperatorTok{!}\NormalTok{server\_is}\OperatorTok{(}\StringTok{"auto"}\OperatorTok{))} \OperatorTok{\{}
\NormalTok{      this}\OperatorTok{{-}\textgreater{}}\NormalTok{connectHost }\OperatorTok{=}\NormalTok{ SuplConfig}\OperatorTok{::}\NormalTok{Instance}\OperatorTok{(){-}\textgreater{}}\NormalTok{server}\OperatorTok{;}      \CommentTok{// e.g. supl.google.com}
\NormalTok{      this}\OperatorTok{{-}\textgreater{}}\NormalTok{connectPort }\OperatorTok{=}\NormalTok{ SuplConfig}\OperatorTok{::}\NormalTok{Instance}\OperatorTok{(){-}\textgreater{}}\NormalTok{port}\OperatorTok{;}        \CommentTok{// e.g. 7275}
      \ControlFlowTok{goto}\NormalTok{ use\_address}\OperatorTok{;}
    \OperatorTok{\}}
  \OperatorTok{\}}
  \CommentTok{// fall back to the standard 3GPP well{-}known H{-}SLP hostname,}
  \CommentTok{// derived from the device\textquotesingle{}s own SIM IMSI (MCC/MNC) {-}{-} not}
  \CommentTok{// attacker{-}influenceable at all}
\NormalTok{  snprintf}\OperatorTok{(}\NormalTok{hostBuf}\OperatorTok{,} \KeywordTok{sizeof}\OperatorTok{(}\NormalTok{hostBuf}\OperatorTok{),}
           \StringTok{"h{-}slp.mnc}\SpecialCharTok{\%03d}\StringTok{.mcc}\SpecialCharTok{\%03d}\StringTok{.pub.3gppnetwork.org"}\OperatorTok{,}\NormalTok{ mnc}\OperatorTok{,}\NormalTok{ mcc}\OperatorTok{);}
\NormalTok{  this}\OperatorTok{{-}\textgreater{}}\NormalTok{connectHost }\OperatorTok{=}\NormalTok{ hostBuf}\OperatorTok{;}
\NormalTok{  this}\OperatorTok{{-}\textgreater{}}\NormalTok{connectPort }\OperatorTok{=}\NormalTok{ SuplConfig}\OperatorTok{::}\NormalTok{Instance}\OperatorTok{(){-}\textgreater{}}\NormalTok{port}\OperatorTok{;}
\NormalTok{use\_address}\OperatorTok{:}
  \OperatorTok{...}
\OperatorTok{\}}
\end{Highlighting}
\end{Shaded}

For a non-emergency SUPL\_INIT (\texttt{isEmergencySuplNi\ ==\ false},
which is what any purely SMS-based attack must send, since it cannot
place the handset into a real emergency call), the packet's own address
(\texttt{this-\textgreater{}eslpHost}) is never consulted at all. The
destination is either the configured server (\texttt{acSuplServer} in
\texttt{gps.xml}, \texttt{supl.google.com} on this device) or, if that
is unset, a standard hostname derived from the device's own SIM
identity. This substantiates conclusion 3 and, for the non-emergency
case, closes off any possibility of an attacker-chosen destination.

\subsection{5. Dynamic analysis}\label{dynamic-analysis}

Dynamic testing had two goals: (a) confirm that a genuine, over-the-air
SMS actually reaches this code, and (b) confirm the static-analysis
conclusions above against the real, running binary rather than
decompiled pseudocode alone.

\subsubsection{5.1 Confirming real SMS
delivery}\label{confirming-real-sms-delivery}

Sending the test SUPL\_INIT produces, in the device's own RIL client
log, delivery of the expected unsolicited message to \texttt{gpsd}'s
registered RIL client:

\begin{verbatim}
RILClient: [OemClient]IND: (clientId = 16, msgId = 4011, dataLength = 259, channel = 0)
\end{verbatim}

\texttt{msgId\ =\ 4011} is \texttt{RILC\_UNSOL\_GPS\_SUPL\_NI\_MESSAGE};
\texttt{clientId\ =\ 16} is \texttt{gpsd}'s own process/thread,
confirming the message reached \texttt{gpsd}'s registered handler rather
than merely the modem/RIL layer.

\subsubsection{5.2 Confirming the permission gate denies under normal
conditions}\label{confirming-the-permission-gate-denies-under-normal-conditions}

With \texttt{gpsd} instrumented to log entry and return value of
\texttt{RildClientHelper::HasPermissionSuplNi} and every function
downstream of it, sending the test (non-emergency) SUPL\_INIT SMS while
the handset was in its normal, idle state (no active call) produced:

\begin{verbatim}
[RildClientHelper::HasPermissionSuplNi] ENTER isEmergencySuplNi=0x0 isInEmergencyState=0x0
[RildClientHelper::HasPermissionSuplNi] RETURNED 0x0   (0 = DENY, 1 = ALLOW)
\end{verbatim}

No hook downstream of this call (session creation, connection requests,
IPC to \texttt{scd}) fired at all. This is the live confirmation of
conclusion 1: under real conditions, the message is processed as far as
this check and then silently discarded.

\subsubsection{5.3 Confirming the destination address for a bypassed,
non-emergency
request}\label{confirming-the-destination-address-for-a-bypassed-non-emergency-request}

To verify Section 4.5's destination-resolution logic against the actual
running code, the permission gate was forced to return ``allow''
(\texttt{Interceptor.attach} return-value override on
\texttt{HasPermissionSuplNi}) so that the connection logic downstream
could be observed operating on our still-non-emergency test payload.
This is not something an SMS attacker can do; it isolates and directly
tests the \texttt{autoConfigSlp} logic identified statically in Section
4.5.

\begin{verbatim}
[RildClientHelper::HasPermissionSuplNi] ENTER isEmergencySuplNi=0x0 isInEmergencyState=0x0
[PATCH] HasPermissionSuplNi forced to 1 (ALLOW)
[BrcmGpsHalScdClient::RequestConnectionForNetwork] ENTER
    param2(network)="supl.google.com"  param3(port)=7275
[WRITE fd=23] len=904        <- connection request sent to scd over IPC
[BrcmGpsHalScdClient::OnIpcMessage] type=4 (OnConnect)
[BrcmGpsHalScdClient::OnIpcMessage] type=5 (OnClose)           <- ~6 ms later
\end{verbatim}

Even with the permission gate bypassed, the connection request carries
\texttt{supl.google.com:7275} (the handset's own configured SLP server),
not any address from the attacker's SUPL\_INIT payload. A direct live
read of the running process's configuration singleton confirmed the same
value:

\begin{verbatim}
SuplConfig::Instance()->server = "supl.google.com"
SuplConfig::Instance()->port   = 7275
\end{verbatim}

This is the live confirmation of Section 4.5 and conclusion 3: for a
non-emergency message, the destination is never attacker-controlled,
under any condition tested, including with the permission gate itself
bypassed.

\subsubsection{5.4 Summary of dynamic results against static
predictions}\label{summary-of-dynamic-results-against-static-predictions}

{\def\LTcaptype{none} 
\begin{longtable}[]{@{}
  >{\raggedright\arraybackslash}p{(\linewidth - 2\tabcolsep) * \real{0.5000}}
  >{\raggedright\arraybackslash}p{(\linewidth - 2\tabcolsep) * \real{0.5000}}@{}}
\toprule\noalign{}
\begin{minipage}[b]{\linewidth}\raggedright
Static prediction (Section 4)
\end{minipage} & \begin{minipage}[b]{\linewidth}\raggedright
Dynamic result
\end{minipage} \\
\midrule\noalign{}
\endhead
\bottomrule\noalign{}
\endlastfoot
Non-emergency SUPL\_INIT denied by \texttt{HasPermissionSuplNi} unless
in an active emergency call (4.3) & Confirmed: gate returned
\texttt{DENY} for our test payload in the handset's normal (non-call)
state (5.2) \\
Emergency-only path required for attacker-address use (4.4) &
Consistent: our test payload was never emergency-flagged, and, per 5.3,
its address was never used even with the gate bypassed \\
Non-emergency destination resolves to configured server, not packet
content (4.5) & Confirmed: live connection request and live config read
both showed \texttt{supl.google.com:7275} (5.3) \\
\end{longtable}
}

\subsection{6. Comparison with an iPhone
8}\label{comparison-with-an-iphone-8}

On the iPhone 8 handset studied, SUPL is disabled: the handset ignores
any SUPL\_INIT message it receives and does not open SUPL connections of
its own. We establish this from its code and its live configuration
(Sections 6.3 to 6.6; see the limitations at the end of this section).

We repeated the static analysis on an Apple iPhone 8 and checked the
results against the live state of the handset. The iPhone gates
network-initiated SUPL more tightly than the Pixel 8:

\begin{enumerate}
\def\labelenumi{\arabic{enumi}.}
\tightlist
\item
  \emph{All SUPL policy is in Apple's own code.} The decision whether to
  act on a SUPL\_INIT, and where to connect, is made in Apple's
  \texttt{locationd} daemon, not in modem or GNSS vendor code.
\item
  \emph{A SUPL\_INIT is ignored unless the handset is in an emergency
  call or the carrier explicitly allows otherwise.} As on the Pixel, the
  gate is the handset's own emergency-call state, which an SMS cannot
  set. A carrier can waive it with a bundle flag
  (\texttt{allowNonEmergencySupl}), but no carrier bundle shipped by
  Apple sets that flag.
\item
  \emph{A server address carried in the SUPL\_INIT is used only if it
  matches the carrier's configured server.} Unlike on the Pixel, this
  holds even for an emergency SUPL\_INIT received during an emergency
  call.
\item
  \emph{On the test handset, SUPL is disabled outright.} Its SIM's
  carrier bundle configures no SUPL server, so every SUPL\_INIT is
  refused.
\item
  \emph{The handset does not use SUPL for ordinary A-GPS.} Assistance
  data is a plain file download from Apple that carries no identifier,
  whereas the Pixel makes SUPL requests of its own to
  \texttt{supl.google.com} (Section 1).
\end{enumerate}

\subsubsection{6.1 Setup}\label{setup}

\textbf{Handset.} Apple iPhone 8 (iPhone10,1, model MQ6L2), Qualcomm
baseband (firmware 8.50.04), iOS 16.7.11 (build 20H360), the same
handset and build as in {[}Leith-iOS{]}. Jailbroken with palera1n.
Fitted with a SIM from 48 (MCC 272, MNC 11), an Irish operator that uses
Three Ireland's network.

\textbf{Static analysis.} \texttt{/usr/libexec/locationd} (source
version 2787.0.5) and \texttt{/usr/libexec/gpsd} (207.0.10) were
decompiled with IDA Pro (IDA Home 9.4, Hex-Rays). Unlike the Pixel's
Broadcom daemon, \texttt{locationd} logs extensively and its log strings
include C++ class and method names, so functions and object fields can
be identified from the binary itself. Apple publishes no documentation
on SUPL in iOS, and we are not aware of prior analysis of its handling.

\subsubsection{6.2 Where the checks are
made}\label{where-the-checks-are-made}

The SMS is received by Apple's telephony daemon \texttt{CommCenter} and
passed to \texttt{locationd}, which selects one of two back-ends
according to the hardware model
(\texttt{CLGpsController::createInUniverse}). On Qualcomm-modem handsets
(\texttt{CLGpsController41}) the modem runs the SUPL session and asks
\texttt{locationd} whether to accept each SUPL\_INIT; \texttt{locationd}
replies accept or deny. On newer handsets (\texttt{CLNilrSupl}),
\texttt{locationd} decides and hands an accepted session to Apple's
\texttt{gpsd}, which connects only to the server address
\texttt{locationd} supplies. On the test handset the cached controller
object has the \texttt{CLGpsController41} class (read live), so the
Qualcomm path is the one in use.

\subsubsection{6.3 The permission gate}\label{the-permission-gate-1}

The Qualcomm-path decision is made in
\texttt{CLGpsController41::handleQmiEmergencySuplInit} (address
\texttt{0x100c15cc8}). Decompiled and cleaned up for readability:

\begin{Shaded}
\begin{Highlighting}[]
\DataTypeTok{void}\NormalTok{ CLGpsController41}\OperatorTok{::}\NormalTok{handleQmiEmergencySuplInit}\OperatorTok{(}\DataTypeTok{const}\NormalTok{ SuplInfo }\OperatorTok{\&}\NormalTok{suplInfo}\OperatorTok{)}
\OperatorTok{\{}
  \DataTypeTok{int}\NormalTok{ niResp }\OperatorTok{=} \DecValTok{0}\OperatorTok{;}                                   \CommentTok{// default: deny}
  \ControlFlowTok{if} \OperatorTok{(}\NormalTok{fSuplNiSessionActive}\OperatorTok{)}         \ControlFlowTok{goto}\NormalTok{ done}\OperatorTok{;}      \CommentTok{// a session is already running}
  \ControlFlowTok{if} \OperatorTok{(}\NormalTok{suplInfo}\OperatorTok{.}\NormalTok{hasTriggerType}\OperatorTok{)}      \ControlFlowTok{goto}\NormalTok{ done}\OperatorTok{;}      \CommentTok{// triggered/periodic sessions refused}
  \ControlFlowTok{if} \OperatorTok{(!}\NormalTok{fEmergencySuplSupported}\OperatorTok{)}     \ControlFlowTok{goto}\NormalTok{ done}\OperatorTok{;}      \CommentTok{// "emergency Supl not supported. Deny"}
  \ControlFlowTok{if} \OperatorTok{(!}\NormalTok{isInCellularEmergency}\OperatorTok{(}\NormalTok{fEmergencyState}\OperatorTok{)} \OperatorTok{\&\&}
      \OperatorTok{!(}\NormalTok{fSuplFlags }\OperatorTok{\&}\NormalTok{ allowNonEmergencySupl}\OperatorTok{))}
                                    \ControlFlowTok{goto}\NormalTok{ done}\OperatorTok{;}
\NormalTok{  niResp }\OperatorTok{=}\NormalTok{ slpAddressAcceptable}\OperatorTok{(}\NormalTok{suplInfo}\OperatorTok{.}\NormalTok{slpAddress}\OperatorTok{);}   \CommentTok{// Section 6.4}
\NormalTok{done}\OperatorTok{:}
  \OperatorTok{...}                                               \CommentTok{// decision returned for the modem}
\OperatorTok{\}}
\end{Highlighting}
\end{Shaded}

\texttt{fEmergencySuplSupported} and \texttt{fSuplFlags} come from the
carrier bundle (Section 6.5). The emergency state is pushed to
\texttt{locationd} by its emergency controller
(\texttt{CLEmergencyController}), which takes it from the telephony
stack's emergency-mode notifications and from the system call state
(\texttt{onEmergencyCallStatusChanged}). \texttt{isInCellularEmergency}
requires an emergency call over cellular on the emergency bearer; a text
session with emergency services also counts. Nothing in the SUPL\_INIT
or in any SMS feeds this state. The \texttt{CLNilrSupl} path applies the
same conditions: \texttt{process\_RawSuplInit} drops the message with
``received SUPL init while not in emergency or not allowed in
non-emergnecy. Dropping'' (sic), and also requires SUPL version 2.

\subsubsection{6.4 The server address}\label{the-server-address}

If the SUPL\_INIT carries a server address, it is accepted only when it
is empty (the session then uses the carrier's configured server, the
H-SLP), when the carrier sets \texttt{allowHslpMismatch}, or when it
matches the carrier's H-SLP: exactly, or as a pattern if the carrier
sets \texttt{allowHslpWildCard}. On the \texttt{CLNilrSupl} path, an
address in a SUPL\_INIT without the emergency notification is ignored
unless the carrier sets \texttt{allowSuplWithoutEmergencyNotification},
and the port always comes from the carrier configuration. So, unlike the
Pixel, the iPhone does not connect to an attacker-chosen server even
when the SUPL\_INIT is emergency-flagged and arrives during a genuine
emergency call, except where a wildcard pattern is written loosely
(Section 6.5).

\subsubsection{6.5 Carrier and operator
configuration}\label{carrier-and-operator-configuration}

iOS takes operator-specific settings (access point names, voice-over-LTE
options, emergency-call settings and so on) from bundles: configuration
packages, one per operator, that are signed by Apple and ship with iOS
in \texttt{/System/Library/Carrier\ Bundles/iPhone/}.
\texttt{CommCenter} assigns each SIM two of them: a \emph{carrier
bundle}, chosen from the SIM's own identity (its MCC and MNC and, for
some operators, further SIM fields), and an \emph{operator bundle},
chosen from the network the handset is registered on. Both are drawn
from the same set of bundles. For a SIM used on its own operator's
network the two are the same bundle; they differ when the SIM's operator
uses another operator's network, e.g.~when it is a mobile virtual
network operator (MVNO). Each bundle holds a \texttt{carrier.plist} plus
per-model override files. \texttt{locationd} reads its SUPL settings
from a \texttt{SUPL} dictionary in the bundle
(\texttt{SuplHSLPServerAddress}, \texttt{SuplFlags},
\texttt{SuplCARootCert}, \texttt{SuplUdpPortConfig}), which in iOS
16.7.11 is mostly in the per-model override files and in iOS 27.0.1 is
in \texttt{carrier.plist}. It enables SUPL only if both the carrier
bundle and the operator bundle give an H-SLP and the two are the same.

{\def\LTcaptype{none} 
\begin{longtable}[]{@{}
  >{\raggedright\arraybackslash}p{(\linewidth - 4\tabcolsep) * \real{0.3333}}
  >{\raggedright\arraybackslash}p{(\linewidth - 4\tabcolsep) * \real{0.3333}}
  >{\raggedright\arraybackslash}p{(\linewidth - 4\tabcolsep) * \real{0.3333}}@{}}
\toprule\noalign{}
\begin{minipage}[b]{\linewidth}\raggedright
\end{minipage} & \begin{minipage}[b]{\linewidth}\raggedright
iOS 16.7.11 (handset)
\end{minipage} & \begin{minipage}[b]{\linewidth}\raggedright
iOS 27.0.1 (iPhone17,3 restore image)
\end{minipage} \\
\midrule\noalign{}
\endhead
\bottomrule\noalign{}
\endlastfoot
Bundles (all operators, about 200 countries) & 660 & 682 \\
Bundles with a \texttt{SUPL} dictionary & 26 & 26 \\
Operators with SUPL configured & Verizon and its US partners, KDDI,
SoftBank, Rakuten (Japan), KT, SKT (Korea) & same \\
Irish bundles & Three, eir, O2, Vodafone: no SUPL & the same plus Sky:
no SUPL \\
SUPL server addresses & all emergency servers (\texttt{e-slp.}) &
same \\
\texttt{allowNonEmergencySupl},
\texttt{allowSuplWithoutEmergencyNotification},
\texttt{allowHslpMismatch} & set in no bundle & set in no bundle \\
\texttt{allowHslpWildCard} & Verizon family, Rakuten & Verizon family,
Rakuten \\
\end{longtable}
}

\textbf{Table 1.} SUPL settings in Apple's bundles, which serve as both
carrier and operator bundles. The table shows that (i) only 26 of the
660 bundles configure SUPL at all, all of them for emergency positioning
by five operator groups in three countries, (ii) no bundle relaxes the
emergency gate or the server-address check, and (iii) nothing changed in
these respects between iOS 16.7.11 and iOS 27.0.1.

On the test handset, Apple ships no bundle for the SIM's MCC and MNC
(48, Section 6.1), so its carrier bundle is Apple's fallback for
unrecognised SIMs (\texttt{Unknown.bundle}), while its operator bundle
is Three Ireland's. Neither contains SUPL settings. The values read live
from \texttt{locationd} with Frida 17.18.0 agree: emergency SUPL
supported 0, carrier server address empty, \texttt{suplFlags} 0x0. Every
SUPL\_INIT on this handset is therefore refused at the ``emergency Supl
not supported'' check, before the emergency-call check is reached.

\textbf{Wildcard matching.} Where \texttt{allowHslpWildCard} is set,
\texttt{locationd} compiles the carrier's entry,
e.g.~\texttt{e-slp.lte.911.*.myvzw.com}, as a regular expression and
matches it against the whole host name. Its dots are not escaped and so
match any character. With the same C++ standard library implementation
(LLVM libc++ \texttt{std::regex}, on macOS), we confirmed that
\texttt{e-slp.lte.911.x.evil-myvzw.com} and
\texttt{e-slp.x.sos.supl-rakuten.jp}, domains outside the carriers' own
that anyone could register, are accepted, while appending a suffix
(\texttt{e-slp.lte.911.x.myvzw.com.attacker.net}) is rejected. This
matters only for Verizon-family and Rakuten SIMs, during a genuine
emergency call, and we have not examined whether TLS server
authentication would then reject an attacker's certificate.

\textbf{Updates.} Apple also distributes updated bundles. On the test
handset, \texttt{CommCenter}'s log shows it checking for updates
automatically (every 10 to 23 hours, and when network access becomes
available), verifying the signatures of installed bundles, and set to
install updates without user interaction (``Bundle updates will be
silent''); we did not observe an update being installed. Since updates
are signed by Apple, an operator presumably cannot change these settings
itself, over the network or through the SIM; a SIM update can only
change which of Apple's existing bundles is selected. Through Apple,
though, an operator could have SUPL settings added to its bundle and
installed on handsets without the user's involvement.

\subsubsection{6.6 Handset-generated SUPL}\label{handset-generated-supl}

The iPhone obtains A-GPS assistance without SUPL. \texttt{locationd}
downloads Qualcomm XTRA predicted-orbit files from Apple
(\texttt{https://iphone-ld.apple.com/xtra/xtra3grej.bin} and
\texttt{xtra3int.bin}). Each request is a bare HTTPS GET with no query
string, cookie or authorisation header, and a User-Agent that identifies
only the software version
(\texttt{locationd/2787.0.5\ CFNetwork/1410.1\ Darwin/22.6.0}).

To look for SUPL connections directly, we restarted \texttt{locationd}
twice, 35 minutes apart (killing it with Frida; \texttt{launchd}
restarted it within seconds), while recording the handset's log and
capturing its traffic at the Wi-Fi access point: every new TCP
connection, all DNS lookups, all UDP other than DNS and QUIC, and all
packets on the SUPL ports 7275 to 7277. The SIM has no prepaid credit,
so the handset's data goes over Wi-Fi.

On restart, \texttt{locationd}'s SUPL state machine
(\texttt{CLSuplHelper}) logs that it starts in its disabled state
(\texttt{kStateDisabled}). Each carrier-settings update that follows
sends it a disable event, and it logs ``SUPL: already disabled''; in the
three minutes of log we recorded after the first restart, the state
machine reports no other state and no SUPL session starts. The relevant
log entries are (process name and log level removed, ``\ldots{}'' marks
omitted fields):

\begin{verbatim}
07:02:32.0   (locationd killed; launchd starts a new instance)
07:02:38.445 SUPL: state,kStateDisabled,2,event,kEventEnter,0
07:02:38.488 SUPL: state,kStateDisabled,2,event,kEventDisable,15
07:02:38.490 #Warning SUPL: already disabled
07:02:41.015 SUPL: carrier settings update, ..., newSuplState,<private>, ...
07:02:41.016 SUPL: state,kStateDisabled,2,event,kEventDisable,15
07:02:41.017 #Warning SUPL: already disabled
   ...        (the same three lines after each of six further updates)
07:02:43.426 SUPL: state,kStateDisabled,2,event,kEventBasebandUp,7
\end{verbatim}

The log redacts the settings themselves
(\texttt{\textless{}private\textgreater{}}), so on the second restart we
read them with a Frida hook on the function that logs them,
\texttt{CLGpsController41::handleEmergencySettingsChange}. For all seven
updates, on entry and on return, the previous and new SUPL state are 0,
the H-SLP address is empty, the port is the default 7275 and
\texttt{suplFlags} is 0, as in the live values of Section 6.5.

Over the 37 minutes captured, which cover both restarts, the handset
opened only HTTPS connections, all to Apple addresses, and looked up
only Apple host names (its location servers, including
\texttt{iphone-ld.apple.com} for the XTRA files, plus push,
certificate-status and App Store hosts). Neither restart was followed by
traffic on the SUPL ports or by a lookup of a SUPL server.

\subsubsection{6.7 Comparison}\label{comparison}

{\def\LTcaptype{none} 
\begin{longtable}[]{@{}
  >{\raggedright\arraybackslash}p{(\linewidth - 4\tabcolsep) * \real{0.3333}}
  >{\raggedright\arraybackslash}p{(\linewidth - 4\tabcolsep) * \real{0.3333}}
  >{\raggedright\arraybackslash}p{(\linewidth - 4\tabcolsep) * \real{0.3333}}@{}}
\toprule\noalign{}
\begin{minipage}[b]{\linewidth}\raggedright
\end{minipage} & \begin{minipage}[b]{\linewidth}\raggedright
Pixel 8
\end{minipage} & \begin{minipage}[b]{\linewidth}\raggedright
iPhone 8
\end{minipage} \\
\midrule\noalign{}
\endhead
\bottomrule\noalign{}
\endlastfoot
Where SUPL\_INIT policy is decided & Broadcom GNSS vendor daemon
(\texttt{gpsd}) & Apple \texttt{locationd} \\
Emergency gate & handset's emergency-call state & same, unless the
carrier sets \texttt{allowNonEmergencySupl} (no bundle does) \\
Server address from the SUPL\_INIT, during an emergency & used; the
emergency APN allowlist is empty & used only if it matches the carrier's
server \\
Non-emergency SUPL\_INIT & address ignored; configured server used &
refused \\
SUPL on the test handset & enabled (\texttt{supl.google.com}) & disabled
(no carrier SUPL settings) \\
Handset-generated SUPL & yes, to \texttt{supl.google.com} & not used for
A-GPS; XTRA file download instead \\
Source of configuration & \texttt{gps.xml} on the vendor partition &
Apple-signed carrier bundles \\
\end{longtable}
}

\textbf{Table 2.} SUPL\_INIT handling on the two handsets. The table
shows that both gate on the handset's own emergency-call state, but the
iPhone additionally constrains the destination server and, on the test
handset, has SUPL disabled entirely.

\textbf{Limitations.} This comparison rests on static analysis and on
reading the live configuration; no SUPL\_INIT was delivered to the
iPhone. On the Qualcomm path the SUPL session itself runs in the modem,
whose firmware we did not analyse. Whether the modem could act on a
SUPL\_INIT without \texttt{locationd}'s approval, or open SUPL
connections of its own, is therefore not established. Such connections
would use the modem's own data session, which our Wi-Fi captures cannot
see. These gaps can only mean that we have missed SUPL activity, not
overstated it.

\subsection{7. Use Of AI}\label{use-of-ai}

The Pixel 8 analysis (Sections 3 to 5) was carried out with the
assistance of Claude (Anthropic, Claude Sonnet 5), which helped run the
tests, carry out the static analysis and prepare the report. The iPhone
8 analysis (Section 6) was carried out and drafted by Claude (Anthropic,
Claude Opus 5.5), working in Claude Code under the author's direction.
The author reviewed the report.

\subsection{8. References}\label{references}

\begin{itemize}
\tightlist
\item
  {[}OMA-AD-SUPL{]} Open Mobile Alliance, ``Secure User Plane Location
  Architecture,'' OMA-AD-SUPL-V2\_0-20120417-A, 17 April 2012.
\item
  {[}OMA-TS-ULP{]} Open Mobile Alliance, ``UserPlane Location
  Protocol,'' OMA-TS-ULP-V2\_0\_5-20191028-A, 28 October 2019.
\item
  {[}OMA-TS-ULP-V3{]} Open Mobile Alliance, ``UserPlane Location
  Protocol,'' OMA-TS-ULP-V3\_0-20181213-C, 13 December 2018.
\item
  {[}Leith-iOS{]} D. J. Leith, ``Decoding What iOS Location Services
  Send to Apple: An LLM Is an Unreliable Analyst but a Useful
  Toolmaker,'' arXiv:2610.21597, October 2026.
\end{itemize}

\end{document}